\documentclass[10pt,twoside,reqno]{amsart}
\usepackage{mathbbol,mathtools,slashed}

\usepackage{multirow}
\usepackage[bookmarksnumbered, plainpages]{hyperref}
\usepackage[rightcaption]{sidecap}
\usepackage{caption,xparse}
\usepackage{pgfplots}

\usepackage{float}

\usepackage[english]{babel}

\usepackage[latin1]{inputenc}
\usetikzlibrary{positioning,arrows,patterns}

\usepackage{graphicx}
\usepackage{epstopdf}
\usepackage{subfig}
\usepackage{color}
\usepackage{amsthm}
\usepackage{amsmath, amsfonts, amssymb}
\usepackage[figuresright]{rotating}
\begin{document}
   \title{Real-time thermal self-energies: In the variational bases and spaces}
   \author{{\small M. A. A. AHMED$^{1,2,\dag}$
    H. ZAINUDDIN$^{1,3,\ddag}$, N. M. Shah$^{1,3,\star}$}}
    \thanks{\scriptsize
 emails:\\ $^\dag$ mohammed\_h7@$_{\text{yahoo.com}}^{\text{taiz.edu.ye}}$\\$^\ddag$ hisham@upm.edu.my.\\ $^\star$risya@upm.edu.my}
 \maketitle
\begin{center}
\address{$^{1}$ Laboratory of Computational Sciences and Mathematical Physics, Institute for Mathematical Research, Universiti Putra Malaysia, UPM Serdang 43000, Selangor, Malaysia\\
 $^{2}$Physics Department, Faculty of Science, Taiz University Al-Turba branch, Taiz, Yemen.\\
 $^{4}$ Department of Physics, Faculty of Science, Universiti Putra Malaysia, 43400 Serdang, Selangor, Malaysia.}
\end{center}
\begin{abstract}
In this work, we introduce a systematic study for studying the scalar propagator and tadpole self-energy by considering an arbitrary parameter $\sigma$ that allows for a path integral description in real-time formalism (RTF). The closed time path formalism (CTP) and Thermofield Dynamics (TFD) are two popular choices for the parameter $\sigma$ in the Feynman rules. We have constructed a scalar propagator and a tadpole self-energy in two different bases in the momentum space as well as the mixed space. The results show that the diagonal components of self-energy in both spaces for the 1/2 basis are the same in both approaches within RTF, whereas the other off-diagonal components of self-energy are different because they depend on the path parameter. On the other hand, the diagonal components of self-energy in both spaces for the new basis are not the same in both approaches within RTF, whereas the off-diagonal components of self-energy are vanishing. That means the new basis allows one to reduce the components for the quantities studied, like self-energy or other.
\end{abstract}
\markboth{\rightline {\sl M. A. A. Ahmed et al. }}
        {\leftline{\sl  RTF thermal self-energies: In the variational bases and spaces.}}

\section{Introduction}\label{INTRODUCTION}
To describe a Thermal Field Theory (TFT), there are some approaches, e.g. the Imaginary Time Formalism (ITF) and Real-Time Formalism (RTF). The ITF is useful to study the system in equilibrium. However, the latter is more comfortable than the former in the computations for equilibrium and non-equilibrium cases. The RTF has been widely used to study condensed matter phenomena, phase transitions, solve many-body problems, etc. Within RTF, there are two commonly formalisms, the Schwinger-Keldysh formalism (also known as the closed time path formalism (CTP)) \cite{Schwinger11,Keldysh9}, and the operator formalism so-called ThermoField Dynamics (TFD), was suggested by Takahashi and Umezawa \cite{Takahashi12}. Both approaches are tailor-made for calculating the Green functions \cite{Lundberg2021}. While the complete theory can be reconstructed from Green functions in principle, for practical and theoretical reasons, an operator formulation of TFT may be useful. In particular, it could illustrate why it is important to double the degrees of freedom for the real-time. Within this framework, quantum field theory can naturally be extended to finite temperature.

In RTF, the number of independent fields doubles. Instead of a single scalar field $\phi$, we encounter two scalar fields, $\phi_1$ and $\phi_2$, called type-1 and type-2 fields, respectively. The propagator becomes then a $2\times2$ matrix. In other words, there are two types of vertices, where each type has only fields that emerge from it and has its usual value, e.g. for the four-particle vertex, fields of type-1 are not mixed with fields of type-2. The vertex involving fields of type-2 has a relative minus sign coming from the anti-time ordering.
In this regard, the (11)-component of the propagator consists of the sum of a vacuum part and a finite-temperature part and the integration is taken over a real, continuous, energy $p_0$ due to the presence of the delta function. The finite temperature contribution is trivial to obtain, thus suggesting the possibility of a ``real time'' perturbation theory \cite{KAPUSTA}. It is interesting that perturbation theory at finite temperature can be formulated directly in both the RTF and ITF. The RTF will be adopted in this work.
\begin{figure}[H]
  \center
   \subfloat{{\includegraphics[width=350pt]{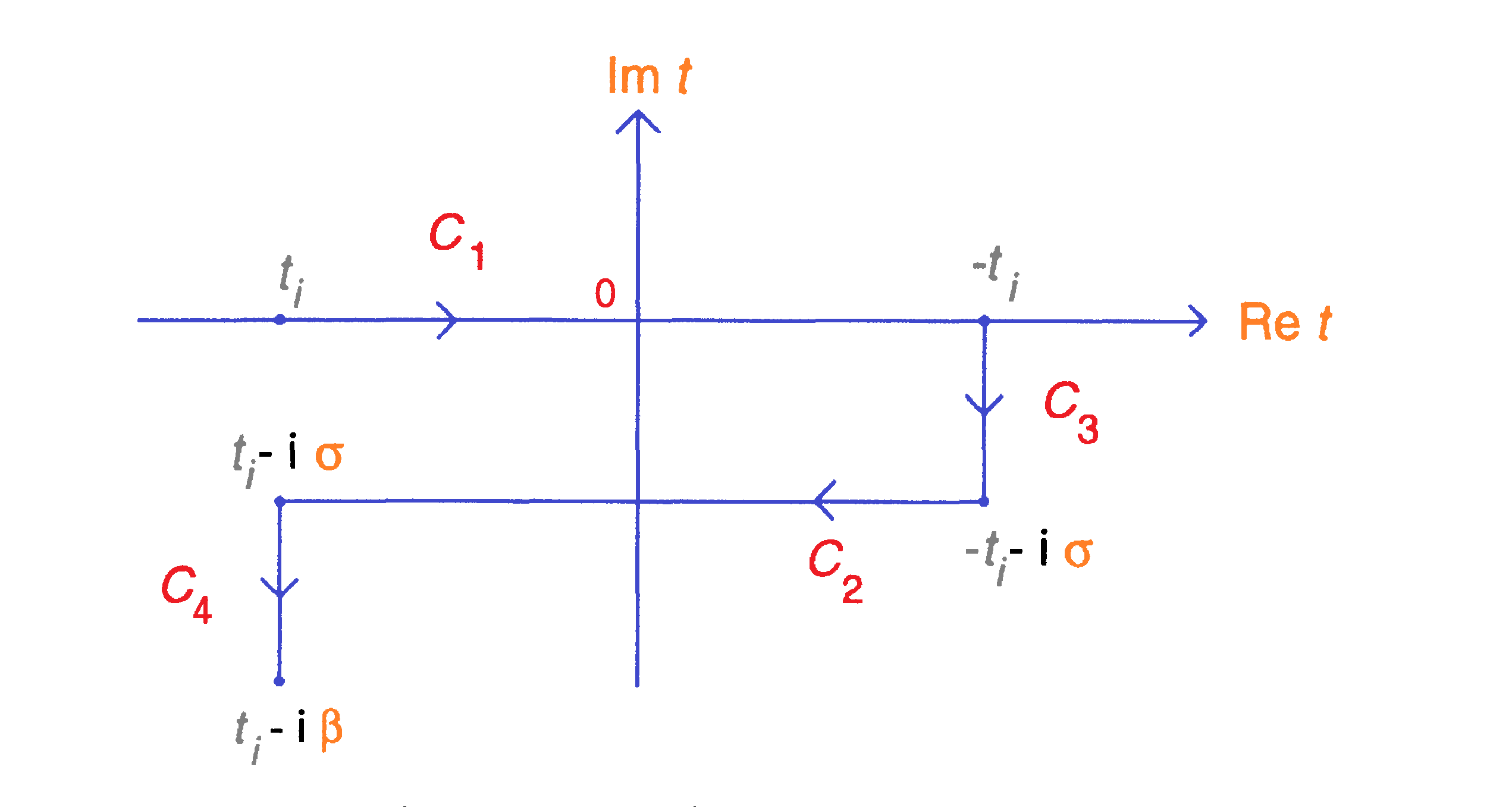}} }

  \caption{The diagram Real-time path contour in the complex time plane.} \label{fig21}
\end{figure}
In most cases, in our work, we need to evaluate causal Green's functions in an arbitrary path for a scalar field (see fig.(\ref{fig21})). In RTF the field degrees of freedom are doubled and the standard Green's function is replaced by four Green's functions (propagators) with arguments on different places of the two parts of the corresponding integration contour. Therefore, the propagators will be written as $2\times2$ matrices whose elements correspond to the usual thermal propagators in the mixed space $(t,\textbf{p})$ \cite{das2006,das2018}.
\begin{equation}\label{propagator}
G^{(\sigma,1/2)}(t,\textbf{p};\beta)=\int_{-\infty}^{\infty}\frac{dq_0}{2\pi}\exp^{-iq_0t}G^{(\sigma,1/2)}(p_0,\textbf{p};\beta),
\end{equation}
where $G^{(\sigma,1/2)}$ is the general propagator with $D^{(\sigma,1/2)}$  stands for the boson propagator or $S^{(\sigma,1/2)}$ the fermion propagator, the coordinate in the momentum space is $(q_0,q)$ and the coordinate in the mixed space is $(t,q)$. Written as $2\times 2$ matrix, it is
 \begin{equation}\label{pr0}
   G^{(\sigma,1/2)}(P;\beta)=\left(\begin{array}{ll}
G^{(\sigma,11)}(P;\beta)&G^{(\sigma,12)}(P;\beta)\\
G^{(\sigma,21)}(P;\beta)&G^{(\sigma,22)}(P;\beta)\end{array}\right),
\end{equation}
with $(0\leq \sigma\leq1)$, $T$ represents temperature and $\beta$ denotes the inverse temperature. Note that for particular values of $\sigma$, there is a one-parameter family of paths in the complex $t$ plane \cite{Matsumoto1983,DAS}. When $\sigma=0$, the real-time description is known as the CTP formalism \cite{Schwinger11,Keldysh9}, while $\sigma=\frac{1}{2}$ leads to TFD \cite{Takahashi12,Khanna2}. A close comparison between the CTP and TFD is presented. We find that TFD and CTP are in many ways the same in form. In particular, the two approaches are identical in stationary situations. However, TFD and CTP are quite different in time-dependent out of equilibrium situations. The main source of this difference is that the time evolution of the density matrix itself is ignored in CTP while in TFD it is replaced by a time-dependent Bogoliubov transformation (BT).
The subscripts $(1,2)$ refer to the two real branches of the time contour in fig (\ref{fig21}); these are the two fields in the $1/2$ basis. So, we have two types of propagator components. The off-diagonal components ($``12"$ and $``21"$), are unphysical because one of the time arguments has an imaginary component. The only physical propagators are the diagonal components ($``11"$ and $``22"$). In addition, we can introduce a second basis, which is very useful for the CTP. It is built from linear combinations of the propagator matrix components of the RTF \cite{Umezawa1994,ahmed2022}. The new basis is obtained from the original 1/2 basis. In other words, in the finite-dimensional case, a change of basis can be represented by a rotation of the axes in the Hilbert space. The two scalar fields will be $\widetilde{\phi_1}$ and $\widetilde{\phi_2}$, and the thermal propagator can be written in the new basis, which involves four independent components. To this end, we make an orthogonal transformation \cite{Smilga1997}.
The aim of the present article is the computation of the scalar propagator and tadpole self-energies in scalar QED. However, we want to treat them in a systematic and comprehensive way when one considers these propagator components in different bases \cite{Ghiglieria2020,ahmed2021a}. Thereafter we use those components to compute the self-energies in the momentum space as well as the mixed space for the general path within RTF \cite{das2018,ahmed2021a}.

 This work is organized as follows. In the next section, the elements of the scalar propagator in the momentum space are given in RTF, defined on any one-parameter $\sigma$ family of paths, where there are two popular choices for the parameter $\sigma$ in the Feynman rules. When $\sigma=0$, the path is CTP and when $\sigma=\frac{1}{2}$, the path is in TFD. We will consider Fourier transform to build the elements of the propagator in the mixed space for the general path, together with CTP and TFD paths. In section \ref{sec2}, we introduce a very convenient representation of the RTF, the new representation obtained via the change of basis. The new basis is obtained from the original basis to represent the propagator elements in the momentum space and the mixed space. In sections \ref{sec3} and \ref{sec4}, we consider the tadpole diagram as an example to introduce the self-energies which were studied in the momentum space and the mixed space for both bases within RTF. Finally, section \ref{sec5} summarizes the results and possibilities for future work.
 \section{\textbf{The scalar propagator on an arbitrary path within RTF in the 1/2 basis}}\label{sec1}
Here, we want to emphasize, as we have mentioned earlier, that in the literature, the two real branches of the path were usually labeled as (1,2) in fig.(\ref{fig21}) for any non-trivial value of $\sigma$. Although the propagator components are different for the various values of $\sigma$, in thermal equilibrium they lead to equivalent physical results for any $\sigma$.
\subsection{The scalar propagator in the momentum space}\label{sec11}
Fig. (\ref{fig21}) refers to the fields moving along $\mathcal{C}_1$. The path follows a horizontal straight line for $\mathcal{C}_{(1,2)}$; $\mathcal{C}_1$ going from $(t_i)$ to $(-t_i)$, $\mathcal{C}_2$ going from $(-t_i-i\sigma)$ to $(t_i-i\sigma)$. Then one goes back to a negative real part moving vertically along $\mathcal{C}_3$ arriving at $(t_i -i\sigma)$. Finally along $\mathcal{C}_4$ we end up with $(t_i -i\sigma)$. The scalar propagator in RTF given in (\ref{pr0}) is for a general path, whose components in the momentum space at finite temperature are parameterised by $(0\leq\sigma\leq1)$, having elements in $1/2$ basis as
\begin{equation}\label{propath}
 \begin{array}{ll}
D^{(\sigma,11)}(Q;\beta)=-i\Delta(Q)-2i\pi n_B(|q_{0}|;\beta)\delta(Q^2),\\

D^{(\sigma,12)}(Q;\beta)=-2i\pi \left(\Theta(-q_0)+n_B(|q_{0}|;\beta)\right)\delta(Q^2)e^{\sigma\beta q_{0}},\\

D^{(\sigma,21)}(Q;\beta)=-2i\pi \left(\Theta(q_0)+n_B(|q_{0}|;\beta)\right)\delta(Q^2)e^{-\sigma\beta q_{0}},\\

D^{(\sigma,22)}(Q;\beta)=-i\Delta^*(Q)-2i\pi n_B(|q_{0}|;\beta)\delta(Q^2).
\end{array}
 \end{equation}
The Feynman propagator is denoted by $\Delta(Q)$, and it's complex conjugate is $\Delta^*(Q)$ and they are defined at zero temperature for a scalar particle as
 $$\Delta(Q)=i(Q^2+i\eta)^{-1}.$$
By virtue of the limit representation of the $\delta$-function,
\begin{equation}\label{del1}
  \delta(Q^2)=\lim_{\eta\rightarrow0}\frac{1}{\pi}\frac{\eta}{(Q^2)^2+\eta^2},
\end{equation}
we can obtain,
\begin{equation}\label{del2}
  \begin{array}{l}
    \Delta(Q)+\Delta^*(Q)=2i\pi \delta(Q^2), \\
    \Delta(Q)-\Delta^*(Q)=\mathcal{P}\frac{1}{Q^2},
  \end{array}
\end{equation}
where $\mathcal{P}\frac{1}{Q^2}$ is the Cauchy principal value, $Q^2=q_0^2-q^2$ the incoming momentum, and $n_B(Q;\beta)$ denotes the Bose-Einstein distribution \cite{Gozzi2011,das2016,Blasone2018}. It is worth noting that the $\sigma$ explicitly appears only in the off-diagonal components of the matrix propagator. The symmetric propagator is the sum of the diagonal components of (\ref{pr0}) and from (\ref{propath}) gives,
\begin{equation}\label{sym12promom}
  \boxed{D^{(\sigma,S)}(Q;\beta)=Tr\left(D^{(\sigma,1/2)}(Q;\beta)\right)=\frac{2\pi}{i}\delta(Q^2)\left[1+2 n_B(|q_{0}|;\beta)\right]}.
\end{equation}
In (\ref{propath}) the propagator within the thermal field theories  can be given a path integral description \cite{das2016}. However, TFD, corresponding to $\sigma=\frac{1}{2}$, also has an operator description in (\ref{propath}), which leads to the symmetric choice in off-diagonal components $D^{\left(\sigma=\frac{1}{2},12\right)}=D^{\left(\sigma=\frac{1}{2},21\right)}$.
 \begin{table}[h]
\center
\caption{The components of the propagator for CTP and TFD formalisms in the momentum space.}

\begin{tabular}{||l|c|c||}
  \hline
   &$\sigma=0$ &$\sigma=\frac{1}{2}$ \\
  \hline
  \hline
  $D^{(\sigma,11)}(Q;\beta)$ &$-i\Delta(Q)-2i\pi n_B(|q_{0}|;\beta)\delta(Q^2)$ & $-i\Delta(Q)-2i\pi n_B(|q_{0}|;\beta)\delta(Q^2)$ \\
 \hline
  $D^{(\sigma,12)}(Q;\beta)$ &$-2i\pi \left(\Theta(-q_0)+n_B(|q_{0}|;\beta)\right)\delta(Q^2)$ & $-2i\pi\delta(Q^2)\frac{e^{\frac{|q_{0}|\beta}{2}}}{e^{\beta |q_{0}|}-1}$\\
  \hline
  $D^{(\sigma,21)}(Q;\beta)$ &$-2i\pi \left(\Theta(q_0)+n_B(|q_{0}|;\beta)\right)\delta(Q^2)$ & $-2i\pi\delta(Q^2)\frac{e^{\frac{|q_{0}|\beta}{2}}}{e^{\beta |q_{0}|}-1}$\\
  \hline
  $D^{(\sigma,22)}(Q;\beta)$ & $-i\Delta^*(Q)-2i\pi n_B(|q_{0}|;\beta)\delta(Q^2)$& $-i\Delta^*(Q)-2i\pi n_B(|q_{0}|;\beta)\delta(Q^2)$\\

\hline
  \hline
  \end{tabular}

\label{tab:a2}
\end{table}
 We note the function $e^{\frac{\beta |q_{0}|}{2}}$ is still taken into account, which appears in off-diagonal components. In CTP the propagator  has the simplification that the temperature dependent part of every component is the same. The correlation of the general path with the closed path will be defined in the relation as
 \begin{equation*}
 \begin{array}{ll}
\boxed{D^{(\sigma,12)}(Q;\beta)=e^{\sigma\beta q_{0}}D^{(\sigma=0,12)}(Q;\beta), \ \ \ \ \text{and} \ \ \ \  D^{(\sigma,21)}(Q;\beta)=e^{-\sigma\beta q_{0}}D^{(\sigma=0,21)}(Q;\beta),}
\end{array}
\end{equation*}
where $e^{\pm\sigma\beta q_{0}}$ is the thermal evolution function, which appears only in off-diagonal components of (\ref{propath}).
\subsection{The scalar propagator in the mixed space}\label{sec12}
In real time, the thermal scalar propagator in the mixed space is related to the zero temperature propagator \cite{das2018,das2005}. Nonetheless, the integration of the propagator is introduced for two spaces over $q_0$ as
\begin{equation}\label{momspace}
\begin{array}{ll}
  \int_{-\infty}^{\infty} D^{(\sigma,1/2)}(q_0,q;\beta)\ \frac{dq_0}{2\pi} \ \ \ \ \ \ \ \text{the \ momentum\ space},\\
  \int_{-\infty}^{\infty} D^{(\sigma,1/2)}(q_0,q;\beta)\  e^{-i q_0t}\frac{dq_0}{2\pi}\ \ \ \ \ \ \ \text{the\ mixed\ space}.
\end{array}
\end{equation}
Then we use (\ref{momspace}) to integrate (\ref{propath}) (see \ref{a}). The propagator turns out to be as follows:
\begin{itemize}
  \item In the momentum space,
\begin{equation}\label{promom0}
 \begin{array}{ll}
D^{(\sigma,11)}(q;\beta)=\frac{1}{iq}\left[\int_{-\infty}^{\infty}q \Delta(Q)\frac{dq_0}{2\pi}+n_B(q;\beta)\right],\\

D^{(\sigma,12)}(q;\beta)=\frac{1}{iq}\left[\frac{1}{2}\left(\Theta(q)e^{-\sigma\beta q}+\Theta(-q)e^{\sigma\beta q}\right)+n_B(q;\beta)\cosh\sigma\beta q\right],\\

D^{(\sigma,21)}(q;\beta)=\frac{1}{iq}\left[\frac{1}{2}\left(\Theta(-q)e^{\sigma\beta q}+\Theta(q)e^{-\sigma\beta q}\right)+n_B(q;\beta)\cosh\sigma\beta q\right],\\

D^{(\sigma,22)}(q;\beta)=\frac{1}{iq}\left[\int_{-\infty}^{\infty}q \Delta^*(Q)\frac{dq_0}{2\pi}+n_B(q;\beta)\right].\\
\end{array}
 \end{equation}
  \item In the mixed space,
 \begin{equation}\label{promix0}
 \begin{array}{ll}
D^{(\sigma,11)}(t,q;\beta)=\frac{1}{iq}\left[n_B(q;\beta)\cos qt+ \frac{1}{2}\left(\Theta(t)e^{-iqt}+\Theta(-t)e^{iqt}\right)\right],\\

D^{(\sigma,12)}(t,q;\beta)=\frac{1}{iq}\left(\frac{1}{2}e^{iq(t+i\sigma\beta)}+n_B(q;\beta)\cos q(t+i\sigma\beta)\right),\\

D^{(\sigma,21)}(t,q;\beta)=\frac{1}{iq}\left(\frac{1}{2}e^{-iq(t-i\sigma\beta)}+n_B(q;\beta)\cos q(t-i\sigma\beta)\right),\\

D^{(\sigma,22)}(t,q;\beta)=\frac{1}{iq}\left[n_B(q;\beta)\cos qt+ \frac{1}{2}\left(\Theta(t)e^{iqt}+\Theta(-t)e^{-iqt}\right)\right].\\
\end{array}
\end{equation}
\end{itemize}
 \begin{table}[h]
\center
\caption{The components of the propagator for CTP and TFD formalisms in the mixed space.}

\begin{tabular}{||l|c|c||}
  \hline
   &$\sigma=0$ &$\sigma=\frac{1}{2}$ \\
  \hline
  \hline
  $D^{(\sigma,11)}(t,q;\beta)$ &$\begin{array}{c}
                                  \frac{1}{iq}\bigl[n_B(q;\beta)\cos qt\\
       + \frac{1}{2}\left(\Theta(t)e^{-iqt}+\Theta(-t)e^{iqt}\right)\bigr]
                                \end{array}$
   & $\begin{array}{c}
       \frac{1}{iq}\bigl[n_B(q;\beta)\cos qt\\
       + \frac{1}{2}\left(\Theta(t)e^{-iqt}+\Theta(-t)e^{iqt}\right)\bigr]
     \end{array}$
    \\
 \hline
  $D^{(\sigma,12)}(t,q;\beta)$ &$\frac{1}{iq}\left(\frac{1}{2}e^{iqt}+n_B(q;\beta)\cos qt\right)$ & $\frac{\cos qt}{iq}e^{\frac{\beta q}{2}}n_B(q;\beta)$\\
  \hline
  $D^{(\sigma,21)}(t,q;\beta)$ &$\frac{1}{iq}\left(\frac{1}{2}e^{-iqt}+n_B(q;\beta)\cos qt\right)$ & $\frac{\cos qt}{iq}e^{\frac{\beta q}{2}}n_B(q;\beta)$\\
  \hline
  $D^{(\sigma,22)}(t,q;\beta)$ & $\begin{array}{c}
  \frac{1}{iq}\bigl[n_B(q;\beta)\cos qt\\
  + \frac{1}{2}\left(\Theta(t)e^{iqt}+\Theta(-t)e^{-iqt}\right)\bigr]
                                 \end{array}$
  & $\begin{array}{c}
      \frac{1}{iq}\bigl[n_B(q;\beta)\cos qt\\
      + \frac{1}{2}\left(\Theta(t)e^{iqt}+\Theta(-t)e^{-iqt}\right)\bigr]
    \end{array}$
  \\

\hline
  \hline
  \end{tabular}

\label{tab:a3}
\end{table}
All elements in (\ref{promix0}) were, introduced in \cite{das2018}. The symmetric propagator in the mixed space of 1/2 basis is,
\begin{equation}\label{sym12promix}
\boxed{D^{(\sigma,S)}(t,q;\beta)=Tr\left(D^{(\sigma,1/2)}(t,q;\beta)\right)=\frac{\cos qt}{iq}\left[1+2 n_B(q;\beta)\right]}.
\end{equation}
When we integrate (\ref{sym12promom}) in the mixed space the result will be similar to (\ref{sym12promix}).
\section{The scalar propagator on an arbitrary path within RTF in the new basis}\label{sec2}
 In this section we introduce a very convenient new representation of the RTF, obtained via change of basis. It is constructed from linear combinations of the components of the RTF Green functions \cite{chou1985,Umezawa1994,Smilga1997}.
\subsection{The scalar propagator in the momentum space}\label{sec21}
One can present the thermal propagator in the new basis which involves each component, it is convenient for many applications to choose a basis. To this end, we make an orthogonal transformation \cite{Smilga1997,Ghiglieria2020}, explicitly as follows:
\begin{equation}\label{tran1}
  \widetilde{D}^{(\sigma,1/2)}(Q;\beta)=\mathbb{T(\theta)}D^{(\sigma,1/2)}(Q;\beta)\mathbb{T(\theta)}^{-1}.
\end{equation}
The transformation matrix mediating between both representations is defined via:
$$\mathbb{T(\theta)}=\left(
      \begin{array}{cc}
        \cos\theta & -\sin\theta \\
        \sin\theta & \cos\theta \\
      \end{array}
    \right).$$
The new basis is obtained from the original 1/2 basis by a rotation of the axes in the space. We distinguish between $\mathbb{T(\theta)}$ and $\mathbb{T(\theta)}^{t}$ in the following to emphasize that $\mathbb{T(\theta)}$ provides a mapping from physical to diagonal coordinates, while $\mathbb{T(\theta)}^{-1}=\mathbb{T(\theta)}^{t}$ provides inverse mapping. Indeed, the matrix elements in this basis are built after we use some algebra to get the four components of the propagator satisfying the causality relation and the orthogonal transformation. Equation (\ref{tran1}) brings the propagator in this form,
$$\widetilde{D}^{(\sigma,1/2)}=\left(
      \begin{array}{cc}
       \widetilde{D}^{(\sigma,11)} & \widetilde{D}^{(\sigma,12)} \\
       \widetilde{D}^{(\sigma,21)} & \widetilde
       {D}^{(\sigma,22)} \\
      \end{array}
    \right).$$
The elements of the propagator in the general path can be expanded in the new basis, while the parameter $\sigma$ in the 1/2 basis will appear in the off-diagonal elements only. After a few algebraic steps, the components of the propagator in the new basis reads
\begin{equation}\label{proARmom}
  \begin{array}{l}
\widetilde{D}^{(\sigma,11)}=D^{(\sigma,11)}+\sin\theta \left[\sin\theta(D^{(\sigma,22)}-D^{(\sigma,11)}) -\cos\theta (D^{(\sigma,12)}+D^{(\sigma,21)})\right],\\
\widetilde{D}^{(\sigma,12)}=D^{(\sigma,12)}-\sin\theta \left[\sin\theta (D^{(\sigma,12)}+D^{(\sigma,21)})-\cos\theta(D^{(\sigma,11)}-D^{(\sigma,22)})\right],\\
\widetilde{D}^{(\sigma,21)}=D^{(\sigma,21)}-\sin\theta \left[\sin\theta (D^{(\sigma,12)}+D^{(\sigma,21)})-\cos\theta(D^{(\sigma,11)}-D^{(\sigma,22)})\right],\\
\widetilde{D}^{(\sigma,22)}=D^{(\sigma,22)}+\sin\theta \left[\sin\theta(D^{(\sigma,11)}-D^{(\sigma,22)})+\cos \theta (D^{(\sigma,12)}+D^{(\sigma,21)})\right].
\end{array}
\end{equation}
The angle $\theta$ is known as Keldysh's angle when it is equal to $\frac{\pi}{4}$. The new components of the propagator in the momentum space produced by inserting the elements of (\ref{propath}) into (\ref{proARmom}), are
\begin{equation}\label{rraa1}
  \begin{array}{l}
\widetilde{D}^{(\sigma,11)}(Q;\beta)=\frac{\pi\delta(Q^2)}{i}\left[1-\left(\Theta(-q_0)e^{\sigma\beta q_0}+\Theta(q_0)e^{-\sigma\beta q_0}\right)+2n_B(Q;\beta)\left[1-\cosh\sigma\beta q_0\right]\right],\\

\widetilde{D}^{(\sigma,12)}(Q;\beta)=\frac{1}{2}\mathcal{P}\frac{1}{Q^2}+\frac{\pi\delta(Q^2)}{i}\left[\left(\Theta(-q_0)e^{\sigma\beta q_0}-\Theta(q_0)e^{-\sigma\beta q_0}\right)+2n_B(Q;\beta)\sinh\sigma\beta q_0\right],\\

\widetilde{D}^{(\sigma,21)}(Q;\beta)=\frac{1}{2}\mathcal{P}\frac{1}{Q^2}-\frac{\pi\delta(Q^2)}{i}\left[\left(\Theta(-q_0)e^{\sigma\beta q_0}-\Theta(q_0)e^{-\sigma\beta q_0}\right)+2n_B(Q;\beta)\sinh\sigma\beta q_0\right],\\

\widetilde{D}^{(\sigma,22)}(Q;\beta)=\frac{\pi\delta(Q^2)}{i}\left[1+\left(\Theta(-q_0)e^{\sigma\beta q_0}+\Theta(q_0)e^{-\sigma\beta q_0}\right)+2n_B(Q;\beta)\left[1+\cosh\sigma\beta q_0\right]\right].
\end{array}
\end{equation}
 In this basis the $\sigma$ explicitly appears for each component of the propagator. The symmetric propagator in this basis is,
\begin{equation}\label{spramom}
\boxed{\widetilde{D}^{(\sigma,S)}(Q;\beta)=Tr\left(\widetilde{D}^{(\sigma,1/2)}(Q;\beta)\right)=\frac{2\pi}{i}\delta(Q^2)\left[1+2 n_B(|q_{0}|;\beta)\right]}.
\end{equation}
 \begin{table}[h]
\center
\caption{The components of the propagator for CTP and TFD formalisms in the momentum space.}
\begin{tabular}{||l|c|c||}
  \hline
   &$\sigma=0$ &$\sigma=\frac{1}{2}$ \\
  \hline
  \hline
  $\widetilde{D}^{(\sigma,11)}(Q;\beta)$ &$0$ & $\frac{\pi\delta(Q^2)}{i}\left[1+2n_B(Q;\beta)\left[1-e^{\frac{\beta |q_0|}{2}}\right]\right]$ \\
 \hline
  $\widetilde{D}^{(\sigma,12)}(Q;\beta)$ &$\begin{array}{c}
                                        \frac{1}{2}\left[\mathcal{P}\frac{1}{Q^2}+2i\pi\delta(Q^2)(\Theta(-q_0)-\Theta(q_0))\right]\\
                                        \approx\frac{1}{Q^2+isgn(q_0)\eta}
  \end{array}$
   & $\frac{1}{2}\mathcal{P}\frac{1}{Q^2}$\\
  \hline
  $\widetilde{D}^{(\sigma,21)}(Q;\beta)$ &$\begin{array}{c}
                                        \frac{1}{2}\left[\mathcal{P}\frac{1}{Q^2}+2i\pi\delta(Q^2)(\Theta(q_0)-\Theta(-q_0))\right]\\ \approx\frac{1}{Q^2-isgn(q_0)\eta}
                                      \end{array}$
   & $\frac{1}{2}\mathcal{P}\frac{1}{Q^2}$\\
  \hline
  $\widetilde{D}^{(\sigma,22)}(Q;\beta)$ & $\frac{2\pi\delta(Q^2)}{i}\left[1+2n_B(Q;\beta)\right]$& $\frac{\pi\delta(Q^2)}{i}\left[1+2n_B(Q;\beta)\left[1+e^{\frac{\beta |q_0|}{2}}\right]\right]$\\

\hline
  \hline
  \end{tabular}

\label{tab:a4}
\end{table}
It is clear that the thermal contribution in CTP comes only from the symmetrical components, while the component $\widetilde{D}^{(\sigma=0,11)}$ vanishes because $D^{(\sigma=0,11)}+D^{(\sigma=0,22)}=D^{(\sigma=0,12)}+D^{(\sigma=0,21)}$. Therefore, we can study the self-energy at finite temperature through the symmetric propagator, due to the work's interest in thermal contribution.

\subsection{The scalar propagator in the mixed space}\label{sec22}
From section (\ref{sec12}), we can obtain the propagator components in the mixed space for the new basis by inserting (\ref{promix0}) into (\ref{proARmom}). Otherwise, we can use the definition in (\ref{momspace}) to integrate (\ref{rraa1}) over $q_0$ and thereby obtaining the following:
\begin{itemize}
  \item In the momentum space,
\begin{equation}\label{promom0}
 \begin{array}{ll}
\widetilde{D}^{(\sigma,11)}(q;\beta)=\frac{1}{iq}\left[1-\left(\Theta(q)e^{-\sigma\beta q}+\Theta(-q)e^{\sigma\beta q}\right)+n_B(q;\beta)\left[1-\cosh \sigma\beta q\right]\right],\\

\widetilde{D}^{(\sigma,21)}(q;\beta)=\frac{1}{2}\int_{-\infty}^{\infty}\mathcal{P}\frac{1}{Q^2}\frac{dq_0}{2\pi},\\

\widetilde{D}^{(\sigma,12)}(q;\beta)=\frac{1}{2}\int_{-\infty}^{\infty}\mathcal{P}\frac{1}{Q^2}\frac{dq_0}{2\pi},\\

\widetilde{D}^{(\sigma,22)}(q;\beta)=\frac{1}{iq}\left[1+\left(\Theta(q)e^{-\sigma\beta q}+\Theta(-q)e^{\sigma\beta q}\right)+n_B(q;\beta)\left[1+\cosh \sigma\beta q\right]\right].\\
\end{array}
 \end{equation}
  \item In the mixed space,
 \begin{equation}\label{rraamix1}
  \begin{array}{l}
\widetilde{D}^{(\sigma,11)}(t,q;\beta)=\frac{\cos qt}{2iq}\left[\left(1-e^{-\sigma\beta q}\right)+2n_B(q;\beta)\left[1-\cosh\sigma\beta q\right]\right],\\

\widetilde{D}^{(\sigma,21)}(t,q;\beta)=\frac{\sin qt}{2q}\left[\left(\Theta(-t)-\Theta(t)\right)+e^{-\sigma\beta q}-2n_B(q;\beta)\sinh\sigma\beta q\right],\\

\widetilde{D}^{(\sigma,12)}(t,q;\beta)=-\frac{\sin qt}{2q}\left[\left(\Theta(t)-\Theta(-t)\right)+e^{-\sigma\beta q}-2n_B(q;\beta)\sinh\sigma\beta q\right],\\

\widetilde{D}^{(\sigma,22)}(t,q;\beta)=\frac{\cos qt}{2iq}\left[\left(1+e^{-\sigma\beta q}\right)+2n_B(q;\beta)\left[1+\cosh\sigma\beta q\right]\right].
\end{array}
\end{equation}
\end{itemize}

 \begin{table}[h]
\center
\caption{The components of the propagator for CTP and TFD formalisms in the mixed space.}

\begin{tabular}{||l|c|c||}
  \hline
   &$\sigma=0$ &$\sigma=\frac{1}{2}$ \\
  \hline
  \hline
  $\widetilde{D}^{(\sigma,11)}(t,q;\beta)$ &$0$ & $\frac{\cos qt}{iq}\left[\frac{1}{2}+\left(1-e^{\frac{\beta q}{2}}\right)n_B(q;\beta)\right]$ \\
 \hline
  $\widetilde{D}^{(\sigma,12)}(t,q;\beta)$ &$-\frac{\sin qt}{q}\Theta(-t)$ & $\frac{\sin qt}{2q}\left(\Theta(-t)-\Theta(t)\right)$\\
  \hline
  $\widetilde{D}^{(\sigma,21)}(t,q;\beta)$ &$-\frac{\sin qt}{q}\Theta(t)$ & $\frac{\sin qt}{2q}\left(\Theta(-t)-\Theta(t)\right)$\\
  \hline
  $\widetilde{D}^{(\sigma,22)}(t,q;\beta)$ & $\frac{\cos qt}{iq}\left[1+2n_B(q;\beta)\right]$& $\frac{\cos qt}{iq}\left[\frac{1}{2}+\left(1+e^{\frac{\beta q}{2}}\right)n_B(q;\beta)\right]$\\

\hline
  \hline
  \end{tabular}
\label{tab:a5}
\end{table}
Here, the symmetric propagator is written as
\begin{equation}\label{spramix}
\boxed{\widetilde{D}^{(\sigma,S)}(t,q;\beta)=Tr\left(\widetilde{D}^{(\sigma,1/2)}(t,q;\beta)\right)=\frac{\cos qt}{iq}\left[1+2 n_B(q;\beta)\right]}.
\end{equation}
From the structures of the propagator's terms, there are several things to note in sections (\ref{sec1}) and (\ref{sec2}). Some of these terms are temperature independent, presented at $T=0$, while others are temperature dependent. The diagonal components of the propagator in 1/2 basis are independent of the arbitrary parameter $\sigma$ that characterizes the path in fig(\ref{fig21}). Then, the off-diagonal elements which do depend on $\sigma$, are related simply as both in the momentum space in (\ref{propath}) and in the mixed space in (\ref{promix0}), respectively. Meanwhile, in the new basis all components of the propagator are dependent on the arbitrary parameter $\sigma$ that characterizes the path in the momentum space (\ref{rraa1}) and in the mixed space (\ref{rraamix1}), respectively \footnote[1]{The fermion propagator in RTF can construct in the momentum space \cite{chou1985,Smilga1997}, and also the mixed space \cite{DAS,das2005}. The fermion propagator for (11)-component in momentum space is,
$$iS^{(\sigma,11)}(K;\beta)=(\slashed{K}+m)\left[\frac{1}{K^2-m^2+i\eta}-2\pi \ n_f(K;\beta) \delta(K^2-m^2)\right],$$
where $n_f(K;\beta)$ is Fermi-Dirac distribution, and $E^2=k^2+m^2$. In mixed space the fermion propagator is quite simple, i.e.
$$iS^{(\sigma,11)}(t,k;\beta)=\int_{-\infty}^{\infty} \frac{dk_0}{2\pi}e^{-i k_0t}S^{(\sigma,11)}(k,k_0;\beta)=\frac{1}{2E}\left[A(\Theta(t)-n_f(k;\beta))e^{-iEt}+B(\Theta(-t)-n_f(k;\beta))e^{iEt}\right],$$ with\\
$A=\gamma^0E-\vec{\gamma}.\vec{p}+m,$ \ \ \ \ $B=-\gamma^0E-\vec{\gamma}.\vec{p}+m$, where $\gamma$ is gamma matrix.}.

\section{The self-energies in the 1/2 basis}\label{sec3}
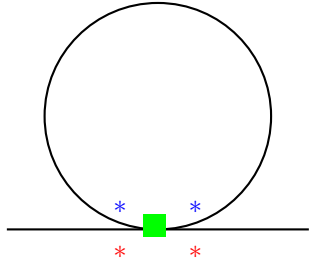
\begin{figure}[h]
\begin{center}
\begin{tikzpicture}
\draw[black,thick](0,0) circle (1.5cm);
  \draw[black,thick](-2,-1.5)--(2,-1.5);
\node at (-0.5,-1.2) {$\textcolor{blue}{*}$};
\node at (0.5,-1.2) {$\textcolor{blue}{*}$};
\node at (-0.5,-1.8) {$\textcolor{red}{*}$};
\node at (0.5,-1.8) {$\textcolor{red}{*}$};

\fill[green] (-0.2,-1.3) rectangle (0.1,-1.6);

\end{tikzpicture}
\end{center}
    \caption{Tadpole diagram: The green blob denotes the effective vertex.}\label{tadpole}
\end{figure}

Now we consider the tadpole diagram of the $\phi^4$-theory shown in fig(\ref{tadpole}) as an example to introduce the self energies. We can see that moving forward in time along the branch $\mathcal{C}_1$ generates a perturbation theory with the same vertices as the usual QFT; we'll call these type-1 vertices. Instead, because the integration along $\mathcal{C}_2$ is backward in time, it produces a set of type-2 vertices with signs that are the opposite of the usual ones. In the RTF it is given as
$$\Pi=\frac{i}{2}(-i\ 4!\ g^2)i\ \int_{0}^{\infty}\frac{d^4Q}{(2\pi)^4}D(Q;\beta),$$
where the symmetry factors are $\frac{1}{2}$ for tadpole, $(-i\ 4!\ g^2)$ is the vertex in vacuum, and $g^2$ is the coupling constant.
In the real time formalism the fields on the two branches of the contour are written as two components of a thermal doublet of scalar fields. Then the general matrix of the tadpole self energies is given by,\\
\begin{equation}\label{ref4}
  \Pi^{(\sigma,1/2)}=\left(\begin{array}{ll}
\Pi^{(\sigma,11)}&\Pi^{(\sigma,12)}\\
\Pi^{(\sigma,21)}&\Pi^{(\sigma,22)}
\end{array}\right),
\end{equation}

where $\Pi$ stands for the self-energy of a boson or fermion. It is obvious that there are two kinds of vertices in the theory for both types of fields, and the non-diagonal matrix structure of (\ref{ref4}) which means that when interactions are included, both types of vertices actually contribute to a given observable. Now let's discuss the two indices, in the 1/2 basis. The tadpole self energies are assigned the same index labelling as the propagators. Therefore we have $\Pi^{(\sigma,11)}$, $\Pi^{(\sigma,12)}$, $\Pi^{(\sigma,21)}$ and $\Pi^{(\sigma,22)}$. The tadpole diagonal components in (\ref{ref4}) can be written like $\Pi^{(\sigma,11)}=-\Pi^{(\sigma,22)}$, because of the minus sign from the ``type-2-field'' vertex of the actions (``type-2 field'' emerging from it and has a value opposite in sign to its type-1 counterpart), and the components $\Pi^{(\sigma,12)}=\Pi^{(\sigma,21)}=0$, since all legs of the vertex must have the same index \cite{Thoma2000}. On the other hand, in the new basis, the thermal effect vanishes for off-diagonal components for self-energy, while thermal effect still exists in the diagonal components.

On the way to calculating the tadpole self energies in the generalized path in RTF, using the components of propagator in the 1/2 basis directly for each component of the self energies, then we introduce them when the value of $\sigma$ corresponding to CTP and TFD formalisms. We next compute the tadpole self energies in a straightforward manner in the new basis.
\subsection{The self-energies for scalar tadpole in the momentum space}\label{sec31}
From the previous explicit expressions of the propagators, we will consider the thermal contribution for self energies of the scalar tadpole as an example. For the use of the following,
\begin{equation}\label{star}
\Pi^{(\sigma,1/2)}=i12g^2\int_{0}^{\infty}\frac{d^4Q}{(2\pi)^4}D^{(\sigma,1/2)}(Q;\beta),
\end{equation}
and trivially integrating over the angles in (\ref{star}), will reduce it to
\begin{equation}\label{selfdemom}
  \Pi^{(\sigma,1/2)}=6\left(\frac{g}{\pi}\right)^2\int_{0}^{\infty}dq\ q^2 \ i \int_{-\infty}^{\infty} \frac{dq_0}{2\pi}D^{(\sigma,1/2)}(Q;\beta).
\end{equation}
We carry out the $q_0$ integral of (\ref{selfdemom}) in the 1/2 basis, then one gets the self energies in the momentum space as,
\begin{equation}\label{self12mom}
\begin{array}{ll}
\Pi^{(\sigma,(11,22))}=6\left(\frac{g}{\pi}\right)^2\ T^2 \zeta(2),\\
\Pi^{(\sigma,(12,12))}=3\left(\frac{g}{\pi}\right)^2\ T^2 \left[\zeta(2,1-\sigma)+\zeta(2,1+\sigma)\right],
\end{array}
 \end{equation}
 \begin{figure}[h]
  \centering
 \includegraphics[width=350pt]{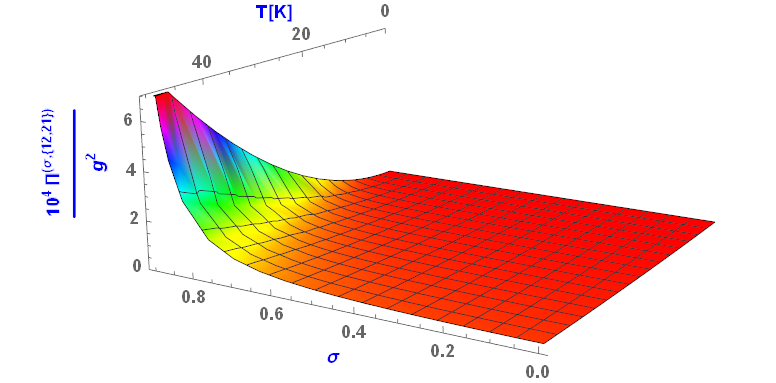}
  \caption{The tadpole self energies $\Pi^{(\sigma,(12,21))}$ in the momentum space.}\label{self12}
\end{figure}
where $\zeta$ is Hurwit Zeta function. The results in (\ref{self12mom}) are calculated only at finite temperature. The symmetric self-energy of tadpole in fig (\ref{tadpole}) will consider,
\begin{equation}\label{se12mom}
\boxed{\Pi^{(\sigma,S)}=Tr\left(\Pi^{(\sigma,1/2)}\right)=3\left(\frac{2g\ T}{\pi}\right)^2\zeta(2)}.
\end{equation}
In TFD, the tadpole self energies of $\Pi^{(\sigma=\frac{1}{2},(11,22))}$, are the same for those which resulted from general path. The tadpole self energies off-diagonal components in TFD are,
$$\Pi^{(\sigma=\frac{1}{2},(12,21))}=6\left(\frac{g}{\pi}\right)^2\ T^2 \zeta\left(2,\frac{1}{2}\right).$$
In CTP, all tadpole self energy components will be the same,
$$\Pi^{(\sigma=0,(11,22,12,21))}=6\left(\frac{g}{\pi}\right)^2\ T^2 \zeta(2).$$
\begin{figure}[h]
  \centering
 \includegraphics[width=170pt]{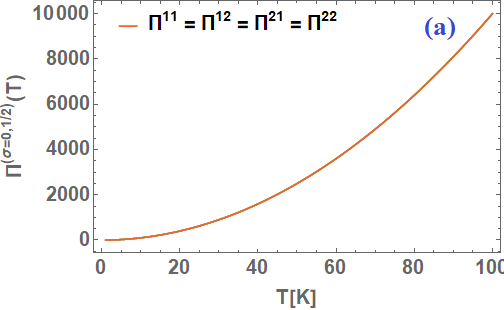}
  \includegraphics[width=170pt]{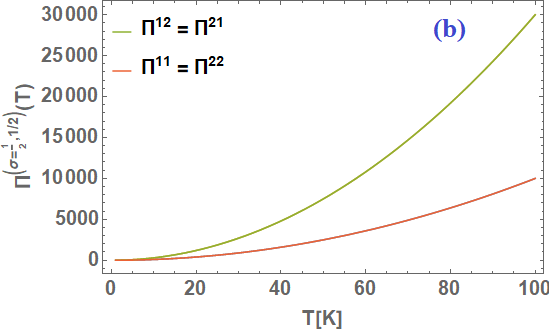}
  \caption{The tadpole self energies in the momentum space. (a) $\Pi^{(\sigma=0,(11,12,21,22))}$ for CTP. (b) $\Pi^{(\sigma=\frac{1}{2},(11,12,21,22))}$ for TFD.}\label{self1}
\end{figure}
We observe the self-energy components of $\Pi^{(\sigma=0,(11,22))}$ are the same for those which resulted from general path and TFD. Obviously, the results in (\ref{self12mom}) for general path will consider of the one component for tadpole as we have mentioned before (see fig (\ref{tadpole})). Nevertheless, if the diagrams have more than one loop, the minus sign from the ``type-2-field'' vertex, which implies that the rules of matrix multiplication do not hold \cite{Thoma2000}. In this study, we assume $g=1$, so the results in the fig (\ref{self1}) show the tadpole self energies of $\Pi^{(11,22)}$, and the parameter $\sigma$ for these components disappears. While the fig (\ref{self12}) illustrates the self energies of $\Pi^{(\sigma,(12,21))}$, and it is clear to see that the increase of these components is dependent on the value of $\sigma$.
\subsection{The self-energies for scalar tadpole in the mixed space}\label{sec32}
As we have mentioned the self energies are related to the bare and full propagators. Here are the results for the self-energy in the mixed space:
\begin{equation}\label{selfdmix}
\Pi^{(\sigma)}_{mix}=6\left(\frac{g}{\pi}\right)^2\int_{0}^{\infty}dq\ q^2 \ i D^{(\sigma)}(t,q;\beta),
\end{equation}
we carry out the $q$ integral of (\ref{selfdmix}) in the 1/2 basis, then one gets the tadpole self energies in the mixed space as
\begin{equation}\label{self12mix}
\begin{array}{ll}
\Pi_{mix}^{(\sigma,(11,22))}=3\left(\frac{g}{\pi}\right)^2\ T^2 \left[-\frac{1}{t^2}+\psi^1(\bar{\tau})+\psi^1(\tau)\right],\\
\Pi_{mix}^{(\sigma,12)}=3\left(\frac{g}{\pi}\right)^2\ T^2 \left[\left(\sigma\beta-it\right)^{-2}+\psi^1(\bar{\tau}+\sigma)+\psi^1(\tau-\sigma)\right],\\
\Pi_{mix}^{(\sigma,21)}=3\left(\frac{g}{\pi}\right)^2\ T^2 \left[\left(\sigma\beta+i\ t\right)^{-2}+\psi^1(\bar{\tau}-\sigma)+\psi^1(\tau+\sigma)\right],
\end{array}
 \end{equation}
where $\psi^n(x)$ is the Polygamma function, and $\tau=\left(1+i\frac{t}{\beta}\right)$ is a complex function on $(t;\beta)$, $\bar{\tau}$ is the conjugate function \cite{Matone2006,Dittrich2019}.\\
The symmetric self energy of the tadpole in the mixed space is
\begin{equation}\label{seramin}
\boxed{\Pi_{mix}^{(\sigma,S)}=Tr\left(\Pi^{(\sigma,1/2)}\right)=6\left(\frac{g\ T}{\pi}\right)^2\left[\psi^1\left(\tau\right)+\psi^1\left(\bar{\tau}\right)\right]}.
\end{equation}
At this point, we show the result in the mixed space for tadpole self-energy in TFD:
\begin{equation}\label{self12mixTFD}
\begin{array}{ll}
\Pi_{mix}^{(\sigma=\frac{1}{2},(11,22))}=3\left(\frac{g}{\pi}\right)^2\ T^2 \left[-\frac{1}{t^2}+\psi^1(\bar{\tau})+\psi^1(\tau)\right],\\
\Pi_{mix}^{(\sigma=\frac{1}{2},(12,21))}=3\left(\frac{g}{\pi}\right)^2\ T^2 \left[\psi^1\left(\bar{\tau}-\frac{1}{2}\right)+\psi^1\left(\tau-\frac{1}{2}\right)\right].\\
\end{array}
 \end{equation}
The tadpole self energies $\Pi_{mix}^{(\sigma=\frac{1}{2},(11,22))}$, are the same with those resulted from the general path. At the same time, we find that all self energies of the tadpole are the same in CTP.
$$\Pi_{mix}^{(\sigma=0,(11,22,12,21))}=3\left(\frac{g}{\pi}\right)^2\ T^2 \left[-\frac{1}{t^2}+\psi^1(\bar{\tau})+\psi^1(\tau)\right].$$
Note that, the tadpole self energies $\Pi_{mix}^{(\sigma=0,(11,22))}$, are the same with those resulted from general path and TFD.
\section{The self energies in the new basis }\label{sec4}
A change of basis can be represented by a rotation of the axes in the space, then the Feynman diagrams can also be represented in the new basis. The diagonal components are the symmetric self-energies. Meanwhile, the self-energies are related to the bare and full propagators. Therefore, the self-energies are written via:
$$\widetilde{\Pi}^{(\sigma,1/2)}=\left(
      \begin{array}{cc}
        \widetilde{\Pi}^{(\sigma,11)} & \widetilde{\Pi}^{(\sigma,12)} \\
        \widetilde{\Pi}^{(\sigma,21)} &  \widetilde{\Pi}^{(\sigma,22)} \\
      \end{array}
    \right).$$
\subsection{The self-energies for scalar tadpole in the momentum space}\label{sec41}
Now, we consider the scalar tadpole as an example using the new basis. We carry out the $q_0$ integral of (\ref{selfdemom}) in the new basis, then one gets the self energies in the momentum space as
\begin{equation}\label{selfRAmom1}
  \begin{array}{l}
\widetilde{\Pi}^{(\sigma,11)}=3\left(\frac{g}{\pi}\right)^2\ T^2 \left[(i \sigma)^{-2}+2\zeta(2)-\left(\zeta(2,1-\sigma)+\zeta(2,1+\sigma)\right)\right],\\

\widetilde{\Pi}^{(\sigma,12)}=\widetilde{\Pi}^{(\sigma,21)}=0,\\

\widetilde{\Pi}^{(\sigma,22)}=3\left(\frac{g}{\pi}\right)^2\ T^2 \left[(\sigma)^{-2}+2\zeta(2)+\left(\zeta(2,1-\sigma)+\zeta(2,1+\sigma)\right)\right].
\end{array}
\end{equation}
So, we find
$$\widetilde{\Pi}^{(\sigma,11)}\simeq -\widetilde{\Pi}^{(\sigma,22)}$$
\begin{figure}[h]
  \centering
 \includegraphics[width=170pt]{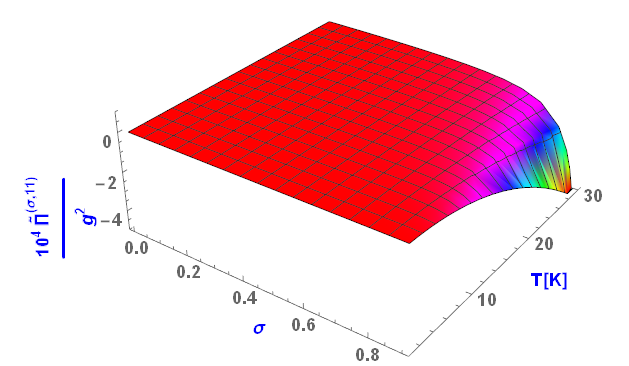}
  \includegraphics[width=170pt]{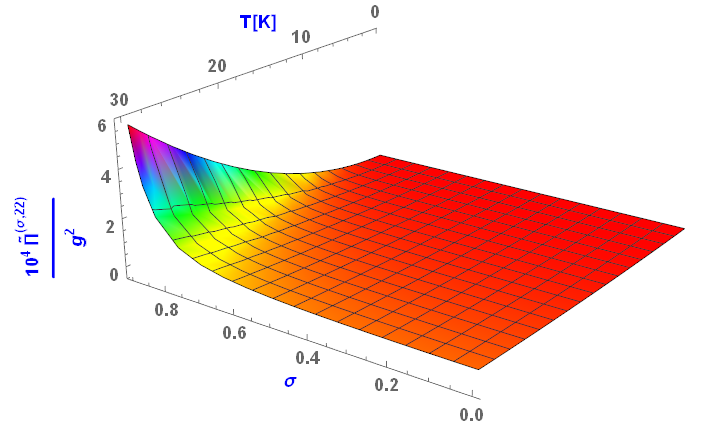}

  \caption{ The tadpole self energies $\widetilde{\Pi}^{(\sigma,(11,22))}$ in the momentum space.}\label{selfra}
\end{figure}
 The symmetric self-energy of tadpole is,
\begin{equation}\label{seramom}
\boxed{\widetilde{\Pi}^{(\sigma,S)}=Tr\left(\widetilde{\Pi}^{(\sigma,1/2)}\right)=3\left(\frac{2g\ T}{\pi}\right)^2\zeta(2)}.
\end{equation}
This confirms the correctness of the results in (\ref{se12mom}) calculated before.
In TFD, the self energies of the tadpole $\widetilde{\Pi}^{(\sigma=\frac{1}{2},(11,22))}$ are,
\begin{equation}\label{selfRAmom3}
  \begin{array}{l}
\widetilde{\Pi}^{(\sigma=\frac{1}{2},11)}=6\left(\frac{g}{\pi}\right)^2\ T^2 \left[\zeta\left(2\right)-\zeta\left(2,\frac{1}{2}\right)\right],\\
\widetilde{\Pi}^{(\sigma=\frac{1}{2},22)}=6\left(\frac{g}{\pi}\right)^2\ T^2 \left[\zeta\left(2\right)+\zeta\left(2,\frac{1}{2}\right)\right].
\end{array}
\end{equation}
Regarding to the off-diagonal components of the self energies, the propagator components of $\widetilde{D}^{(\sigma=\frac{1}{2},(12,21))}$ refer to the non thermal part and hence one cannot study the thermal contribution for $\widetilde{\Pi}^{(\sigma=\frac{1}{2},(12,21))}$.\\
In CTP, the tadpole self energies are,
\begin{equation}\label{selfRAmom2}
  \begin{array}{l}
\widetilde{\Pi}^{(\sigma=0,(11,12,21))}=0,\\
\widetilde{\Pi}^{(\sigma=0,22)}=\frac{12}{\pi^2}(g T)^2\ \zeta(2).
\end{array}
\end{equation}
\begin{figure}[h]
  \centering
 \includegraphics[width=170pt]{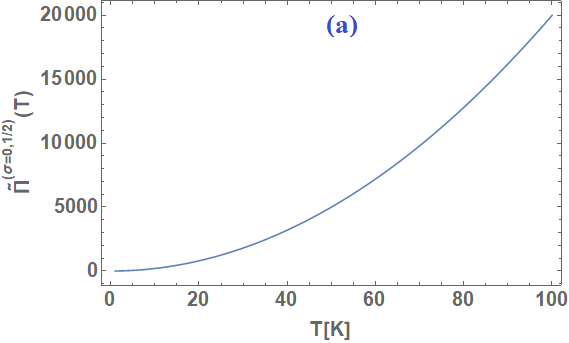}
  \includegraphics[width=170pt]{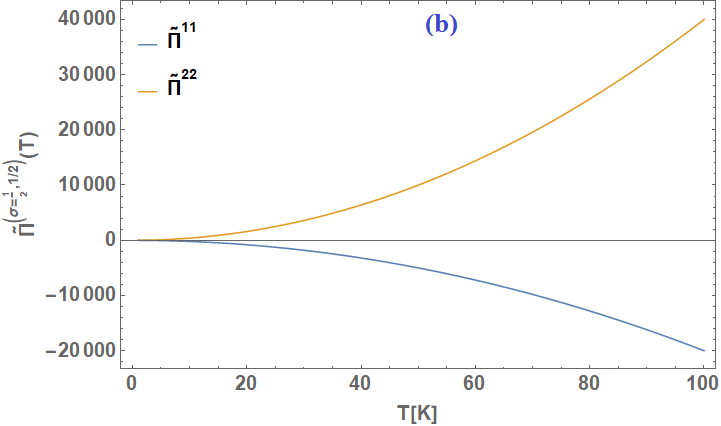}
  \caption{The tadpole self energies in the momentum space. (a) $\widetilde{\Pi}^{(\sigma=0,22)}$ for CTP. (b) $\widetilde{\Pi}^{(\sigma=\frac{1}{2},(11,22))}$ for TFD.}\label{self2}
\end{figure}
The component $\widetilde{D}^{(\sigma=0,11)}$ of propagator is zero in this path, so the self energy $\widetilde{\Pi}^{(\sigma=0,11)}$ equal to zero. While the propagator components for $\widetilde{D}^{(\sigma=0,(12,21))}$ refer to the field at zero temperature only, and thus no thermal effect for $\widetilde{\Pi}^{(\sigma=0,(12,21))}$.
Similar to section (\ref{sec3}), since we know well that all legs of the vertex must have the same index in the tadpole, we considered the two symmetry components $\widetilde{\Pi}^{(\sigma,(11,22))}$ in the new basis. Furthermore, we observed that the symmetric self energies are obtained rather directly in the new basis, whereas the two other components $\widetilde{\Pi}^{(\sigma,(12,21))}$ vanish for both approaches in (\ref{selfRAmom3}) and (\ref{selfRAmom2}). In the meantime, the results show a decrease of the tadpole self energies $\widetilde{\Pi}^{(\sigma,11)}$ in the fig (\ref{selfra}) for the different values of $\sigma$. The reason is the $\zeta$-function has a minus sign in (\ref{selfRAmom1}). While the self energy $\widetilde{\Pi}^{(\sigma,22)}=2(gT)^2$ in fig.(\ref{selfra}) began from the value 2 (when $T=1K$ and $\sigma=0$) it then grows with the path parameter $\sigma$ and temperatures. Fig. \ref{self2} shows us the tadpole self energies in CTP. We have only $\widetilde{\Pi}^{(\sigma=0,22)}$ from (\ref{selfRAmom2}), while the other components are vanishing within this path. The result in (\ref{selfRAmom3}) is for TFD, which manifested the behaviour of the $\widetilde{\Pi}^{(\sigma=\frac{1}{2},22)}$ is similar to itself in CTP. However, component $\widetilde{\Pi}^{(\sigma=\frac{1}{2},11)}$ has different behaviour in the negative region, whereas the $\widetilde{\Pi}^{(\sigma=\frac{1}{2},(12,21))}$ are vanishing as what happened in CTP.
\subsection{The self-energies for scalar tadpole in the mixed space}\label{sec42}
We carry out the $q_0$ integral of (\ref{selfdmix}) in the new basis, then one gets the self energies in the mixed space as
\begin{equation}\label{selfRAmix1}
  \begin{array}{l}
\widetilde{\Pi}_{mix}^{(\sigma,11)}=3\left(\frac{g}{\pi}\right)^2\ T^2 \biggl\{\left[\frac{-(\sigma^2-(t\ T)^2)T^2}{(\sigma^2+(t\ T)^2)^2}\right]\\
+\left[\psi^1(\bar{\tau})+\psi^1(\tau)\right]-\frac{1}{2}\left[\psi^1(\bar{\tau}-\sigma)+\psi^1(\bar{\tau}+\sigma)+\psi^1(\tau-\sigma)+\psi^1(\tau+\sigma)\right]\biggr\},\\

\widetilde{\Pi}_{mix}^{(\sigma,21)}=3\left(\frac{g}{\pi}\right)^2\ T^2 \left[\frac{2i\ t \ T^3}{\left(\sigma^2+(t\ T)^2\right)^{2}}-\frac{1}{2}\left(\psi^1(\bar{\tau}-\sigma)-\psi^1(\bar{\tau}+\sigma)-\psi^1(\tau-\sigma)+\psi^1(\tau+\sigma)\right)\right],\\

\widetilde{\Pi}_{mix}^{(\sigma,12)}=3\left(\frac{g}{\pi}\right)^2\ T^2 \left[\frac{-2i\ t \ T^3}{\left(\sigma^2+(t\ T)^2\right)^{2}}+\frac{1}{2}\left(\psi^1(\bar{\tau}-\sigma)-\psi^1(\bar{\tau}+\sigma)-\psi^1(\tau-\sigma)+\psi^1(\tau+\sigma)\right)\right],\\

\widetilde{\Pi}_{mix}^{(\sigma,22)}=3\left(\frac{g}{\pi}\right)^2\ T^2 \biggl\{\left[\frac{(\sigma^2-(t\ T)^2)T^2}{(\sigma^2+(t\ T)^2)^2}\right]\\
+\left[\psi^1(\bar{\tau})+\psi^1(\tau)\right]+\frac{1}{2}\left[\psi^1(\bar{\tau}-\sigma)+\psi^1(\bar{\tau}+\sigma)+\psi^1(\tau-\sigma)+\psi^1(\tau+\sigma)\right]\biggr\}.\\
\end{array}
\end{equation}
The symmetric self energy of tadpole is,
\begin{equation}\label{seramin}
\boxed{\widetilde{\Pi}_{mix}^{(\sigma,S)}=Tr\left(\Pi^{(\sigma,1/2)}\right)=6\left(\frac{g\ T}{\pi}\right)^2\left[\psi^1\left(\tau\right)+\psi^1\left(\bar{\tau}\right)\right]}.
\end{equation}
The tadpole self-energies in TFD is given by,
\begin{equation}\label{selfRAmix3}
  \begin{array}{l}
\widetilde{\Pi}_{mix}^{(\sigma=\frac{1}{2},11)}=3\left(\frac{g}{\pi}\right)^2\ T^2 \left[\psi^1\left(\bar{\tau}\right)-\psi^1\left(\bar{\tau}-\frac{1}{2}\right)+\psi^1\left(\tau\right)-\psi^1\left(\tau-\frac{1}{2}\right)\right],\\
\\
\widetilde{\Pi}_{mix}^{(\sigma=\frac{1}{2},(12,21))}=0,\\
\\
\widetilde{\Pi}_{mix}^{(\sigma=\frac{1}{2},22)}=3\left(\frac{g}{\pi}\right)^2\ T^2 \left[\psi^1\left(\bar{\tau}\right)+\psi^1\left(\bar{\tau}-\frac{1}{2}\right)+\psi^1\left(\tau\right)+\psi^1\left(\tau-\frac{1}{2}\right)\right].
\end{array}
\end{equation}
The tadpole off-diagonal components of self energy vanish, while the diagonal components remain nonzero.\\
In CTP, the all self energies of tadpole will be,
\begin{equation}\label{selfRAmix2}
  \begin{array}{l}
\widetilde{\Pi}_{mix}^{(\sigma=0,(11,12,21))}=0,\\

\widetilde{\Pi}_{mix}^{(\sigma=0,22)}=6\left(\frac{g}{\pi}\right)^2\ T^2 \left[-\frac{1}{t^2}+\psi^1(\bar{\tau})+\psi^1(\tau)\right].
\end{array}
\end{equation}
Again the results of the $\widetilde{\Pi}_{mix}^{(\sigma=0,(12,21))}$ in CTP are consistent with those results from TFD, and the $\widetilde{\Pi}_{mix}^{(\sigma=0,11)}$ is vanishes within this approach.

\section{Discussion and conclusion }\label{sec5}
In RTF, the fields are doubled accordingly so that the physical field corresponds to the upper component and the auxiliary field corresponds to the lower (virtual) field that only couples to itself. The physical fields, therefore, only connect to auxiliary fields through the off-diagonal thermal propagator. It is worth noting that the $\sigma$ explicitly appears only in the off-diagonal components of the matrix propagator, where $e^{\pm\sigma\beta q_{0}}$ is the thermal evolution function, and it appears in off-diagonal components of (\ref{propath}) in the 1/2 basis. In TFD the thermal evolution function $e^{\frac{\beta |q_{0}|}{2}}$ is still taken into account as shown in table.\ref{tab:a2}, and the off-diagonal components are equivalent to each other, because $D^{12}=D^{21}$ is easily demonstrated \cite{DAS,Gozzi2011,das2016,Blasone2018}. In CTP, the thermal evolution function vanishes in the off-diagonal components of the propagator as shown in table.\ref{tab:a2}. In RTF, however, the diagonal components are $D^{(\sigma,11)}=(D^{(\sigma,22)})^*$ in the general path, TFD, and also CTP.

In the new basis, the parameter $\sigma$ explicitly appears for each component of the propagator in the momentum space (\ref{rraa1}) and the mixed space (\ref{rraamix1}). Meanwhile, the thermal term disappears in off-diagonal components in both approaches of TFD and also CTP, but the diagonal components still exist in TFD, while we only have $\widetilde{D}^{22}$ in CTP. This means that the number of components is reduced in this basis.
In the mixed space, the parameter $\sigma$ appears followed by the time in the off-diagonal components of the propagator in the 1/2 basis while the parameter $\sigma$ appears in each component of the propagator in the new basis, which gives the signal for the flowing of time in the contour.
The symmetric propagator (\ref{sym12promom}) in the 1/2 basis is equivalent to (\ref{spramix}) in the new basis for both spaces. Because the symmetric propagator resulted from the summation of the diagonal components, that exist in the same branch between two points.

Our objective in this work was to analyze the propagator components in two bases and spaces, which are used to construct the tadpole self energies, then we select the diagonal components, due to that all legs of the vertex must have the same index in the tadpole diagram. In other words, the self energies in the new basis correspond to those studied in the 1/2 basis of the ``physical representation'' i.e. $\widetilde{\Pi}^{(\sigma,(11,22))}$ in the new basis and $\Pi^{(\sigma,(11,22))}$ in the 1/2 basis.

In the 1/2 basis, the results in (\ref{self12mom}) and (\ref{self12mix}) are considered only for thermal terms. The $\sigma$ explicitly appears for the off-diagonal components of the self energies in the momentum space and the mixed space, while those components within TFD and also CTP will be different due to path dependence. Whereas the diagonal components of the self energies in the momentum space and the mixed space are similar for the general, TFD and also CTP paths. In addition, the path parameter $\sigma$ for each component of the tadpole self energies in the momentum space (\ref{selfRAmom1}) and the mixed space (\ref{selfRAmix1}) appears explicitly in the new basis. The off-diagonal components vanishes within TFD and also CTP, as well as the $(\widetilde{11})$-component in CTP.

The analysis in the mixed space provides us with the same physical information as the momentum space in principle, while the latter is more comfortable than the former in the computations and gaining physical insights into medium properties. Therefore, we showed the numerical results of tadpole self energies in the momentum space in figs. (\ref{self12}) and (\ref{selfra}), since we have considered only the terms with distribution functions and neglected the other terms.
From the results obtained here, we can conclude that the switch between the two approaches turns out to be related to the $\sigma$ parameter of the time path as used in real time formalism. While the angle between the two bases in this study is $\theta$ for any space.
Besides that, the work was derived in the mixed space, thus the temporal part $e^{-it q_0}$ and the thermal part $e^{\sigma \beta q_0}$ appear, both of them, in the calculations, giving rise to the function depending on the complex time $\tau$ \cite{ahmed2021b}. In the equilibrium situation, the physical system doesn't depend on the time, so we consider the time at zero or near zero. Therefore, the limit of $\tau$ when the time at zero is $\lim _{t\rightarrow 0}\tau=1$, we get the same results in (\ref{momspace}) which obtained when we considered the integration over $q_0$ .
The statistical distribution used with the tadpole in scalar QED is the Bose-Einstein distribution, which plays a key role when treating soft physics. This distribution exists in the thermal part of the propagator for both bases. Therefore, the diagonal-components of the propagator are used for studying the self-energy of a tadpole at finite temperature, as we have mentioned why. A possible future outlook is to consider the thermal propagator of the gluon in QCD to study the tadpole self-energy of the gluon.
\section*{Acknowledgments}
This research was supported by the Fundamental Research Grant Scheme (FRGS) under Ministry of Education with project number FRGS/1/2019/STG02/UPM/02/3.
\appendix
\section{}\label{a}
We introduce the integrations of some quantities:
$$\int_{-\infty}^{\infty}\frac{i}{p_0^2-E^2-i\eta}\frac{dp_0}{2\pi}e^{-ip_0t}=\int_{-\infty}^{\infty}\frac{1}{2E}\left[\frac{i}{p_0+E+i\eta}+\frac{i}{p_0-E-i\eta}\right]\frac{dp_0}{2\pi}e^{-ip_0t},$$
then
 $$\int_{-\infty}^{\infty}\left[\frac{i}{p_0-E-i\eta}\right]\frac{dp_0}{2\pi}e^{-ip_0t}=e^{-itE}\int_{-\infty-E}^{\infty-E}\left[\frac{i}{\chi-i\eta}\right] \frac{d \chi}{2\pi} e^{-i\chi t}=\lim_{\eta \to 0} \Theta(-t)e^{-itE-i \eta}.$$
After integrating over $q_0$ by means of the $\delta$-functions
 $$\int_{-\infty}^{\infty} 2\pi \delta(p_0^2-E^2)n_B(|p_0|;\beta)\frac{dp_0}{2\pi}e^{-ip_0t}=\frac{1}{2E}\int_{-\infty}^{\infty} (\delta(p_0-E)+\delta(p_0+E))n_B(|p_0|;\beta)dp_0 e^{-ip_0t},$$
 \ \ \ \ \ \ \ \ \ \ \ \ \ \ \ \ \ \ \ \ \ \ \ \ \ \ \ \ \ \ \ \ \ \ \ \ \ \ \ \ \ \ \ \ \ \ \ \ \ \ \ \ \ \ \ \ \ \ \ \ \ \ \ \ \ \ \ \ \ \ \ \ \ \ \ \ \ \ \ \ \ \ \ \ \ \ \ \ \ \ \ \ \ $$=\frac{1}{2E}\left[n_B(|E|;\beta)e^{itE}+n_B(|E|;\beta)e^{-itE}\right],$$
 and
 $$\int_{-\infty}^{\infty} \delta(p_0-E)f(p_0;\beta)n_B(|p_0|;\beta)dp_0 e^{-ip_0t}=e^{-itE}f(E;\beta)n_B(|E|;\beta).$$



\section{}\label{b}
In this appendix, we collected all self energies of the tadpole were studied in two bases for both approaches in the momentum space as well as the mixed space, which obtained from the components of the propagator.
\begin{tiny}
\begin{equation*}
\begin{cases}
                   \Pi^{(\sigma,*)}=\begin{cases}
                    \widetilde{\Pi}^{(\sigma,1/2)}= \begin{cases}
                                        \widetilde{\Pi}^{(\sigma=\frac{1}{2},1/2)}=\begin{cases}
                                        \widetilde{\Pi}^{(\sigma=\frac{1}{2},11)}=3\left(\frac{2\ g}{\pi}\right)^2\ T^2 \left[\zeta\left(2\right)-\zeta\left(2,\frac{1}{2}\right)\right],\\
                                         \\
                                         \widetilde{\Pi}^{(\sigma=\frac{1}{2},12)}=0,\\
                                        \\
                                       \widetilde{\Pi}^{(\sigma=\frac{1}{2},21)}=0,\\
                                        \\
                                       \widetilde{\Pi}^{(\sigma=\frac{1}{2},22)}=3\left(\frac{2g}{\pi}\right)^2\ T^2 \left[\zeta\left(2\right)+\zeta\left(2,\frac{1}{2}\right)\right],\\
                                      \end{cases}\\

                                       \widetilde{\Pi}^{(\sigma=0,1/2)}=\begin{cases}
                                       \widetilde{\Pi}^{(\sigma=0,11)}=0,\\
                                                                              \\
                                      \widetilde{\Pi}^{(\sigma=0,12)}=0,\\
                                                                              \\
                                       \widetilde{\Pi}^{(\sigma=0,21)}=0,\\
                                                                              \\
                                      \widetilde{\Pi}^{(\sigma=0,22)}=6\left(\frac{2g}{\pi}\right)^2\ T^2 \zeta(2),\\
                                      \end{cases} \\

                                      \end{cases}
 \\
                \Pi^{(\sigma,1/2)}=     \begin{cases}
                                        \Pi^{(\sigma=\frac{1}{2},1/2)}=\begin{cases}
                                        \Pi^{(\sigma=\frac{1}{2},11)}=6\left(\frac{g}{\pi}\right)^2\ T^2 \zeta(2),\\
                                         \\
                                         \Pi^{(\sigma=\frac{1}{2},12)}=6\left(\frac{g}{\pi}\right)^2\ T^2 \zeta\left(2,\frac{1}{2}\right),\\
                                        \\
                                       \Pi^{(\sigma=\frac{1}{2},21)}=6\left(\frac{g}{\pi}\right)^2\ T^2 \zeta\left(2,\frac{1}{2}\right),\\
                                        \\
                                       \Pi^{(\sigma=\frac{1}{2},22)}=6\left(\frac{g}{\pi}\right)^2\ T^2 \zeta(2),\\
                                      \end{cases}\\

                                       \Pi^{(\sigma=0,1/2)}=\begin{cases}
                                       \Pi^{(\sigma=0,11)}=6\left(\frac{g}{\pi}\right)^2\ T^2 \zeta(2),\\
                                                                              \\
                                      \Pi^{(\sigma=0,12)}=6\left(\frac{g}{\pi}\right)^2\ T^2 \zeta(2),\\
                                                                              \\
                                       \Pi^{(\sigma=0,21)}=6\left(\frac{g}{\pi}\right)^2\ T^2 \zeta(2),\\
                                                                              \\
                                      \Pi^{(\sigma=0,22)}=6\left(\frac{g}{\pi}\right)^2\ T^2 \zeta(2),\\
                                      \end{cases} \\

                                      \end{cases}
 \end{cases}
 \\
                 \Pi_{mix}^{(\sigma,*)} =\begin{cases}
                    \widetilde{\Pi}_{mix}^{(\sigma,1/2)}= \begin{cases}
                                        \widetilde{\Pi}_{mix}^{(\sigma=\frac{1}{2},1/2)}=\begin{cases}
                                        \widetilde{\Pi}_{mix}^{(\sigma=\frac{1}{2},11)}=3\left(\frac{g}{\pi}\right)^2\ T^2 \left[\psi^1\left(\bar{\tau}\right)-\psi^1\left(\bar{\tau}-\frac{1}{2}\right)+\psi^1\left(\tau\right)-\psi^1\left(\tau-\frac{1}{2}\right)\right],\\
                                         \\
                                        \widetilde{ \Pi}_{mix}^{(\sigma=\frac{1}{2},12)}=0,\\
                                        \\
                                      \widetilde{\Pi}_{mix}^{(\sigma=\frac{1}{2},21)}=0,\\
                                        \\
                                       \widetilde{\Pi}_{mix}^{(\sigma=\frac{1}{2},22)}=3\left(\frac{g}{\pi}\right)^2\ T^2 \left[\psi^1\left(\bar{\tau}\right)+\psi^1\left(\bar{\tau}-\frac{1}{2}\right)+\psi^1\left(\tau\right)+\psi^1\left(\tau-\frac{1}{2}\right)\right],\\
                                      \end{cases}\\

                                       \widetilde{\Pi}_{mix}^{(\sigma=0,1/2)}=\begin{cases}
                                       \widetilde{\Pi}_{mix}^{(\sigma=0,11)}=0,\\
                                                                              \\
                                      \widetilde{\Pi}_{mix}^{(\sigma=0,12)}=0,\\
                                                                              \\
                                       \widetilde{\Pi}_{mix}^{(\sigma=0,21)}=0,\\
                                                                              \\
                                      \widetilde{\Pi}_{mix}^{(\sigma=0,22)}=6\left(\frac{g}{\pi}\right)^2\ T^2 \left[-\frac{1}{t^2}+\psi^1(\bar{\tau})+\psi^1(\tau)\right],\\
                                      \end{cases} \\

                                      \end{cases}
 \\
                \Pi^{(\sigma,1/2)}=     \begin{cases}
                                        \Pi_{mix}^{(\sigma=\frac{1}{2},1/2)}=\begin{cases}
                                        \Pi_{mix}^{(\sigma=\frac{1}{2},11)}=3\left(\frac{g}{\pi}\right)^2\ T^2 \left[-\frac{1}{t^2}+\psi^1(\bar{\tau})+\psi^1(\tau)\right],\\
                                         \\
                                       \Pi_{mix}^{(\sigma=\frac{1}{2},12)}=3\left(\frac{g}{\pi}\right)^2\ T^2 \left[\psi^1\left(\bar{\tau}-\frac{1}{2}\right)+\psi^1\left(\tau-\frac{1}{2}\right)\right],\\
                                        \\
                                       \Pi_{mix}^{(\sigma=\frac{1}{2},21)}=3\left(\frac{g}{\pi}\right)^2\ T^2 \left[\psi^1\left(\bar{\tau}-\frac{1}{2}\right)+\psi^1\left(\tau-\frac{1}{2}\right)\right],\\
                                        \\
                                       \Pi_{mix}^{(\sigma=\frac{1}{2},22)}=3\left(\frac{g}{\pi}\right)^2\ T^2 \left[-\frac{1}{t^2}+\psi^1(\bar{\tau})+\psi^1(\tau)\right],\\
                                      \end{cases}\\

                                       \Pi_{mix}^{(\sigma=0,1/2)}=\begin{cases}
                                       \Pi_{mix}^{(\sigma=0,11)}=3\left(\frac{g}{\pi}\right)^2\ T^2 \left[-\frac{1}{t^2}+\psi^1(\bar{\tau})+\psi^1(\tau)\right],\\
                                                                              \\
                                     \Pi_{mix}^{(\sigma=0,12)}=3\left(\frac{g}{\pi}\right)^2\ T^2 \left[-\frac{1}{t^2}+\psi^1(\bar{\tau})+\psi^1(\tau)\right],\\
                                                                              \\
                                       \Pi_{mix}^{(\sigma=0,21)}=3\left(\frac{g}{\pi}\right)^2\ T^2 \left[-\frac{1}{t^2}+\psi^1(\bar{\tau})+\psi^1(\tau)\right],\\
                                                                              \\
                                      \Pi_{mix}^{(\sigma=0,22)}=3\left(\frac{g}{\pi}\right)^2\ T^2                    \left[-\frac{1}{t^2}+\psi^1(\bar{\tau})+\psi^1(\tau)\right],\\
                                      \end{cases} \\

                                      \end{cases}
 \end{cases}
 \\
                 \end{cases}
\end{equation*}
\end{tiny}

\end{document}